\documentclass[aps,prl,preprint,tightenlines,superscriptaddress,showpacs,byrevtex,subfigure]{revtex4}

\usepackage{graphicx}
\usepackage{color}

\begin{document}
\begin{flushright}
 BELLE-CONF-0712\\
\end{flushright}

\title{ \quad\\[0.5cm] 
Search for Lepton Flavor
Violating $\tau$ 
{Decays}\\
into Three Leptons
}
\affiliation{Budker Institute of Nuclear Physics, Novosibirsk}
\affiliation{Chiba University, Chiba}
\affiliation{University of Cincinnati, Cincinnati, Ohio 45221}
\affiliation{Department of Physics, Fu Jen Catholic University, Taipei}
\affiliation{Justus-Liebig-Universit\"at Gie\ss{}en, Gie\ss{}en}
\affiliation{The Graduate University for Advanced Studies, Hayama}
\affiliation{Gyeongsang National University, Chinju}
\affiliation{Hanyang University, Seoul}
\affiliation{University of Hawaii, Honolulu, Hawaii 96822}
\affiliation{High Energy Accelerator Research Organization (KEK), Tsukuba}
\affiliation{Hiroshima Institute of Technology, Hiroshima}
\affiliation{University of Illinois at Urbana-Champaign, Urbana, Illinois 61801}
\affiliation{Institute of High Energy Physics, Chinese Academy of Sciences, Beijing}
\affiliation{Institute of High Energy Physics, Vienna}
\affiliation{Institute of High Energy Physics, Protvino}
\affiliation{Institute for Theoretical and Experimental Physics, Moscow}
\affiliation{J. Stefan Institute, Ljubljana}
\affiliation{Kanagawa University, Yokohama}
\affiliation{Korea University, Seoul}
\affiliation{Kyoto University, Kyoto}
\affiliation{Kyungpook National University, Taegu}
\affiliation{\'Ecole Polytechnique F\'ed\'erale de Lausanne (EPFL), Lausanne}
\affiliation{University of Ljubljana, Ljubljana}
\affiliation{University of Maribor, Maribor}
\affiliation{University of Melbourne, School of Physics, Victoria 3010}
\affiliation{Nagoya University, Nagoya}
\affiliation{Nara Women's University, Nara}
\affiliation{National Central University, Chung-li}
\affiliation{National United University, Miao Li}
\affiliation{Department of Physics, National Taiwan University, Taipei}
\affiliation{H. Niewodniczanski Institute of Nuclear Physics, Krakow}
\affiliation{Nippon Dental University, Niigata}
\affiliation{Niigata University, Niigata}
\affiliation{University of Nova Gorica, Nova Gorica}
\affiliation{Osaka City University, Osaka}
\affiliation{Osaka University, Osaka}
\affiliation{Panjab University, Chandigarh}
\affiliation{Peking University, Beijing}
\affiliation{University of Pittsburgh, Pittsburgh, Pennsylvania 15260}
\affiliation{Princeton University, Princeton, New Jersey 08544}
\affiliation{RIKEN BNL Research Center, Upton, New York 11973}
\affiliation{Saga University, Saga}
\affiliation{University of Science and Technology of China, Hefei}
\affiliation{Seoul National University, Seoul}
\affiliation{Shinshu University, Nagano}
\affiliation{Sungkyunkwan University, Suwon}
\affiliation{University of Sydney, Sydney, New South Wales}
\affiliation{Tata Institute of Fundamental Research, Mumbai}
\affiliation{Toho University, Funabashi}
\affiliation{Tohoku Gakuin University, Tagajo}
\affiliation{Tohoku University, Sendai}
\affiliation{Department of Physics, University of Tokyo, Tokyo}
\affiliation{Tokyo Institute of Technology, Tokyo}
\affiliation{Tokyo Metropolitan University, Tokyo}
\affiliation{Tokyo University of Agriculture and Technology, Tokyo}
\affiliation{Toyama National College of Maritime Technology, Toyama}
\affiliation{Virginia Polytechnic Institute and State University, Blacksburg, Virginia 24061}
\affiliation{Yonsei University, Seoul}
  \author{K.~Abe}\affiliation{High Energy Accelerator Research Organization (KEK), Tsukuba} 
  \author{I.~Adachi}\affiliation{High Energy Accelerator Research Organization (KEK), Tsukuba} 
  \author{H.~Aihara}\affiliation{Department of Physics, University of Tokyo, Tokyo} 
  \author{K.~Arinstein}\affiliation{Budker Institute of Nuclear Physics, Novosibirsk} 
  \author{T.~Aso}\affiliation{Toyama National College of Maritime Technology, Toyama} 
  \author{V.~Aulchenko}\affiliation{Budker Institute of Nuclear Physics, Novosibirsk} 
  \author{T.~Aushev}\affiliation{\'Ecole Polytechnique F\'ed\'erale de Lausanne (EPFL), Lausanne}\affiliation{Institute for Theoretical and Experimental Physics, Moscow} 
  \author{T.~Aziz}\affiliation{Tata Institute of Fundamental Research, Mumbai} 
  \author{S.~Bahinipati}\affiliation{University of Cincinnati, Cincinnati, Ohio 45221} 
  \author{A.~M.~Bakich}\affiliation{University of Sydney, Sydney, New South Wales} 
% ...
\collaboration{The Belle Collaboration}

\begin{abstract}
We search for lepton-flavor-violating $\tau$ decays 
into  three leptons (electron or muon) 
using
535 fb$^{-1}$ of data collected 
with the Belle detector at the 
KEKB asymmetric-energy $e^+e^-$ collider. 
No evidence for these decays is observed and  
we set 90\% confidence level upper limits 
on the branching fractions between 
{$2.0\times 10^{-8}$ and 
$4.1\times 10^{-8}$.}
These results improve the 
{best}
previously published upper limits by factors from 
{4.9 to 7.0.}
\end{abstract}
\pacs{11.30.Fs; 13.35.Dx; 14.60.Fg}
\maketitle
 \section{Introduction}

{Lepton flavor violation (LFV)
{appears}
in {various} extensions of the Standard Model (SM),
{e.g.,} 
{supersymmetry (SUSY), leptoquark and many other models.}}
{In particular, {lepton-flavor-violating} decays
into $\tau^-\to\ell^-\ell^+\ell^-$
(where $\ell = e$ or $\mu$ )
are discussed
{in 
various SUSY models~\cite{cite:susy1,
cite:susy2,cite:susy3,cite:susy4,cite:susy5,
cite:susy6,cite:susy7},
models with
{little Higgs~\cite{cite:littlehiggs1,cite:littlehiggs2},}
{left-right symmetric models~\cite{cite:leftright}}
as well as models with
heavy singlet Dirac neutrinos~\cite{cite:amon}
and
very light pseudoscalar bosons~\cite{cite:pseudo}.}
Some of these models with certain combinations of parameters 
predict that the branching fractions 
for $\tau^-\to\ell^-\ell^+\ell^-$ decays
can be as high as $10^{-7}$, 
which is already accessible 
{in} high-statistics  $B$ factory experiments.
{Searches for LFV $\tau$ decays into three leptons have
a long history~\cite{13}, which starts from the pioneering experiment
of MARKII~\cite{14}.
In previous high-statistics analyses,}
both Belle and BaBar reached 
90\% confidence level (C.L.) upper limits 
on the branching fractions 
{of the} 
order of $10^{-7}$~\cite{cite:3l_belle, cite:3l_babar}, based on 
about 90 fb${}^{-1}$ of data.
Here, we update 
{our} previous results using  535 fb$^{-1}$ of data collected 
with the Belle detector 
{at the KEKB  
asymmetric-energy 
$e^+e^-$ 
collider~\cite{kekb},}
{taken at} 
the $\Upsilon(4S)$ resonance and 60 MeV below it.}

The Belle detector is a large-solid-angle magnetic spectrometer that
consists of a silicon vertex detector (SVD),
a 50-layer central drift chamber (CDC),
an array of aerogel threshold Cherenkov counters (ACC), 
a barrel-like arrangement of
time-of-flight scintillation counters (TOF), 
and an electromagnetic calorimeter
comprised of 
CsI(Tl) {crystals (ECL), all located} inside
a superconducting solenoid coil
that provides a 1.5~T magnetic field.
An iron flux-return located {outside} the coil is instrumented to 
detect $K_{\rm{L}}^0$ mesons
and to identify muons (KLM).
The particle identification is based on the 
ratio of the energy deposit in the ECL to 
the momentum measured in the SVD and CDC, 
the shower shape in the ECL, 
the particle range in the KLM, 
the hit information from the ACC,
the $dE/dx$ information in the CDC, 
and the particle time-of-flight from the TOF.
The detector is described in detail elsewhere~\cite{Belle}.
The leptons are identified 
using likelihood ratio,
(${\cal P}(e)$) for electrons~\cite{EID} 
and  
(${\cal P}({\mu})$) for muons~\cite{MUID}, 
which are based on electron and muon probabilities, respectively.

In order to estimate the signal efficiency and 
to optimize the event selection, 
we use {Monte Carlo} (MC) samples.
The signal and the background events from generic $\tau^+\tau^-$ decays are 
 generated by KORALB/TAUOLA~\cite{KKMC}. 
In the signal MC, we generate $\tau^+\tau^-$, where 
a $\tau$ decays into three leptons and the other $\tau$ decays 
generically.
All leptons from $\tau^-\to\ell^-\ell^+\ell^-$ decays are assumed to have a 
uniform angular distribution in the $\tau$ lepton's 
{rest frame~\cite{XX}.}
Other backgrounds including
$B\bar{B}$ and $e^+e^-\to q\bar{q}$ ($q=u,d,s,c$) events, Bhabha events, $e^+e^-\rightarrow\mu^+\mu^-$, 
and two-photon processes are generated by 
EvtGen~\cite{evtgen},
BHLUMI~\cite{BHLUMI}, 
KKMC~\cite{KKMC}, and
AAFHB~\cite{AAFH}, respectively. 
All kinematic variables are calculated in the laboratory frame
unless otherwise specified.
In particular,
variables
calculated in the $e^+e^-$ center-of-mass (CM) system
are indicated by the superscript ``CM''.

\section{Event Selection}

We search for $\tau^+\tau^-$ events in which one $\tau$ 
decays into three leptons~({signal $\tau$}), 
while the other $\tau$ decays 
into  one charged track, any number of additional 
photons, and neutrinos~({tag $\tau$}).
Candidate $\tau$-pair events are required to have 
four tracks with a zero net charge.
Thus, the decay chain to be reconstructed is:
\begin{center}
{$\left\{
\tau^- \rightarrow \ell^-\ell^+\ell^-
\right\} 
~+
~ \left\{ \tau^+ \rightarrow ({\rm a~track})^+ + (n\geq0~\gamma)
 + X(\rm{missing}) 
\right\}$}\footnotemark[2].  
\end{center}
\footnotetext[2]{Unless otherwise stated, 
charge-conjugate decays are
{implied}
throughout
this paper.}
Here, 
{all possible combinations of three 
leptons in the signal $\tau$ decay,} 
which include
the  
$e^-e^+e^-$,
$\mu^-\mu^+\mu^-$,
$e^-\mu^+\mu^-$,
$\mu^-e^+e^-$,
$\mu^-e^+\mu^-$, and
$e^-\mu^+e^-$ modes, 
are searched for. 
Because the background components 
differ between
signal decay modes, 
the selections described below are optimized 
separately
for each mode.

The event selection starts by reconstructing 
four charged tracks and any number of photons within the fiducial volume
defined by $-0.866 < \cos\theta < 0.956$,
{where $\theta$ is 
the polar angle {relative} to 
the direction opposite to 
that of 
the {incident} $e^+$ beam in 
{the} laboratory frame.}

The transverse momentum ($p_t$) of each charged track
and energy of each photon ($E_{\gamma}$) 
are required to be $p_t> $ 0.1 GeV/$c$ and $E_{\gamma}>0.1$ GeV,
respectively.
The distance of the closest point for each charged track with respect to the interaction point 
is required to be 
within 
$\pm$0.5 cm in the transverse direction 
{and within} 
$\pm$3.0 cm in the longitudinal direction.

%
%
The particles in an event are then separated into two 
hemispheres referred to as the signal and 
tag sides using the plane perpendicular to the thrust axis~\cite{thrust}. 
The tag side contains a charged track
while the signal side contains three charged tracks.
We require all charged tracks on the signal side 
{to be} 
identified as leptons.
The electron (muon)
identification criteria are ${\cal P}(e) > 0.9$ 
(${\cal P}(\mu) > 0.9$) 
{for} the momentum greater than 
0.3 GeV/$c$ (0.6 GeV/$c$).
The electron (muon) identification efficiency
for 
{our selection criteria}
is 91\% (85\%) 
while 
{the probability to misidentify pion 
as 
{electron} (muon)}
is below 0.5\% (2\%).

%
%
To ensure that the missing particles are neutrinos rather
than photons or charged particles that pass  outside the detector acceptance,
we impose additional requirements on the missing 
momentum  vector $\vec{p}_{\rm miss}$,
which 
is
calculated by subtracting the
vector sum of the momenta of all tracks and photons
from the sum of the $e^+$ and $e^-$ beam momenta.
We require that the magnitude of $\vec{p}_{\rm miss}$
be  greater than 0.4 GeV/$c$,
and that its direction point into the fiducial volume of the
detector.

%
%
To reject $q\bar{q}$ background,
we require that the magnitude of thrust ($T$)
be 0.90 $< T  <$ 0.97 
for all modes except for the $\tau^-\to e^-e^+e^-$ mode 
{for 
{which it should be} 0.90 $ < T  <$ 0.96.}
We also require $5.29$ GeV $< E^{\mbox{\rm{\tiny{CM}}}}_{\rm{vis}} < 9.5$ GeV, 
where $E^{\mbox{\rm{\tiny{CM}}}}_{\rm{vis}}$ 
is the total visible energy in the CM system which is defined as 
the sum of the energies of three leptons,
the charged track on the tag side (with a pion mass hypothesis)
and all photon candidates.

\begin{figure}
\begin{center}
       \resizebox{0.4\textwidth}{0.4\textwidth}{\includegraphics
        {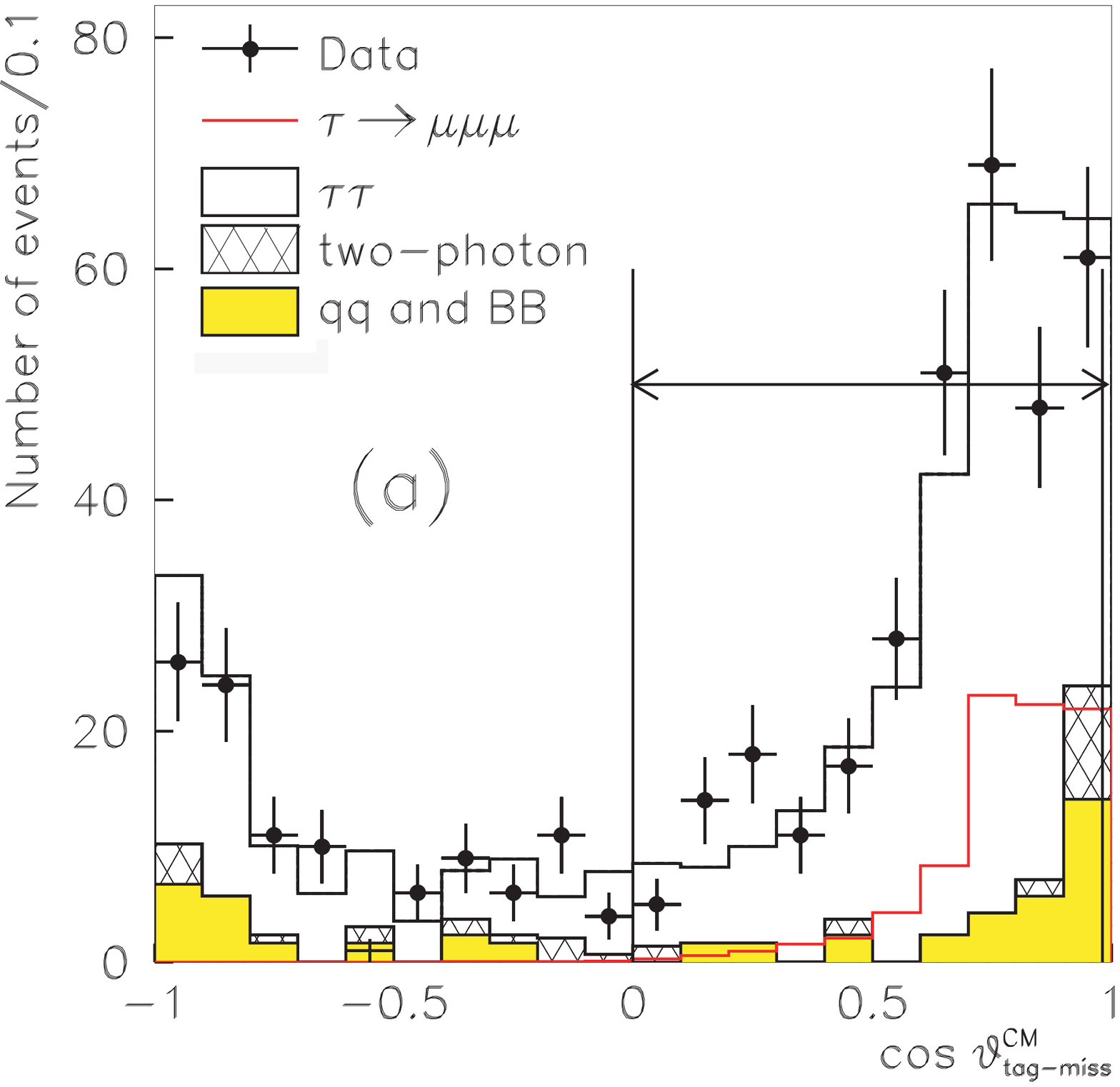}}
       \resizebox{0.4\textwidth}{0.4\textwidth}{\includegraphics
        {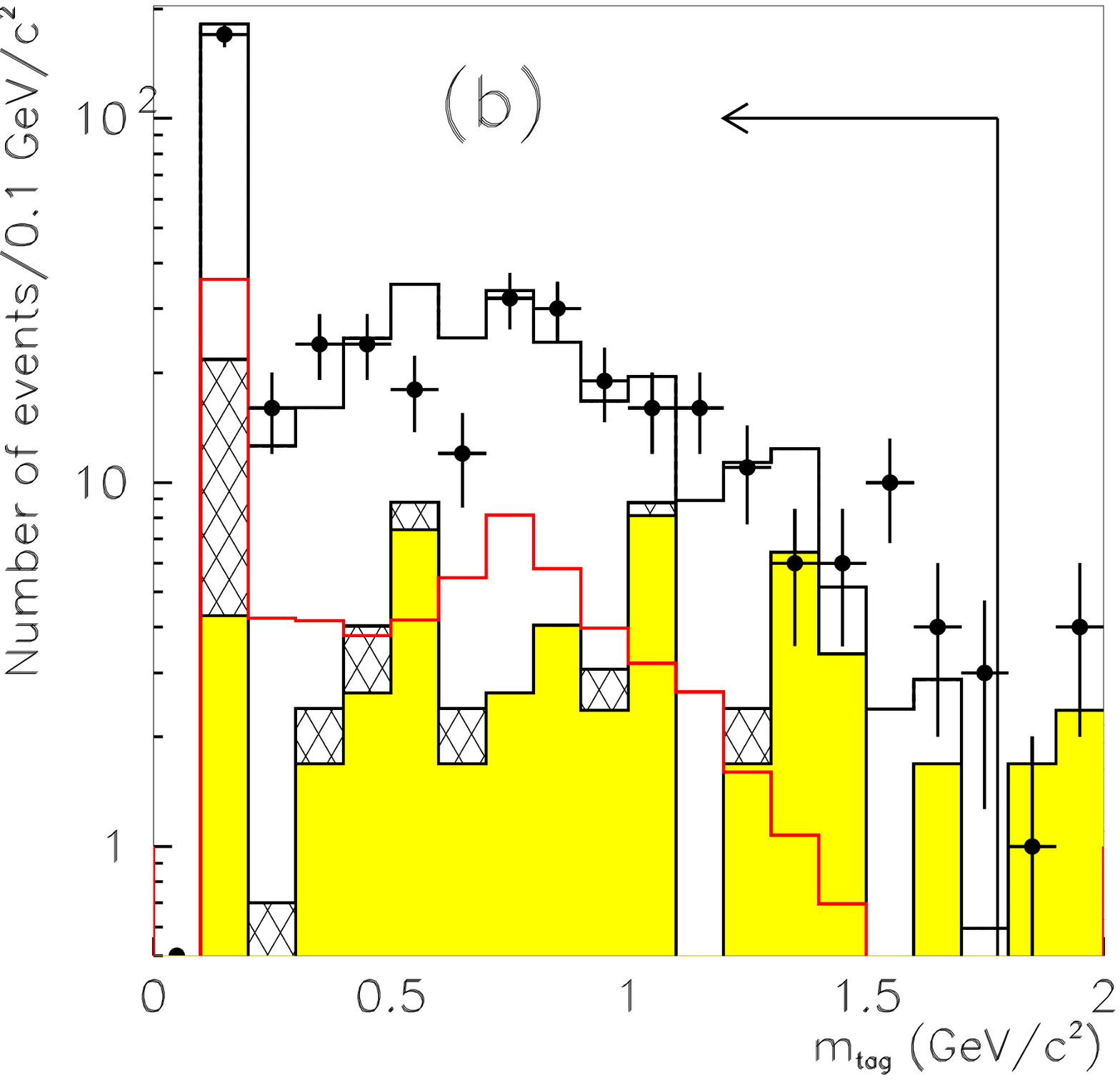}}
 \vspace*{-0.5cm}
 \caption{
 {Kinematic distributions used in the event selection:
 (a) the cosine of the opening angle between a charged track on the 
 tag side and
 missing particles in the CM system ($\cos \theta_{\rm tag-miss}^{\rm CM}$);
 (b) the reconstructed mass on the tag side using a charged track and photons
 after $E^{\rm CM}_{\rm vis}$ and $T$ event selection.
 While the signal MC {($\tau^-\to\mu^-\mu^+\mu^-$)}
 distribution is normalized arbitrarily,
 {the background MC} are normalized to the same luminosity as that of data.
 {Selected regions are indicated
 by {the}
 arrows from the marked cut {boundaries.}}}
}
\label{fig:cut_fig}
\end{center}
\end{figure}

%
%
Since neutrinos are 
emitted only on the tag side,
the direction of
$\vec{p}_{\rm miss}$
should lie within the tag side of the event.
The cosine of the
opening angle between
$\vec{p}_{\rm miss}$
and the charged track on the tag side 
{in the CM system,}
$\cos \theta^{\mbox{\rm \tiny CM}}_{\rm tag-miss}$, 
is therefore required to lie
in the range $0.0<\cos \theta^{\mbox{\rm \tiny CM}}_{\rm tag-miss}<0.98$.
The reconstructed mass on
the tag side using a charged track (with a pion mass hypothesis) and photons,
$m_{\rm tag}$, 
is required to be less than 1.777 GeV/$c^2$.
As shown in Fig.~\ref{fig:cut_fig},
reasonable agreement between data and background MC
is obtained in  the distributions of 
$\cos \theta^{\mbox{\rm \tiny CM}}_{\rm tag-miss}$  
and $m_{\rm tag}$.

%
%

\begin{figure}
 \resizebox{0.37\textwidth}{0.37\textwidth}{\includegraphics
{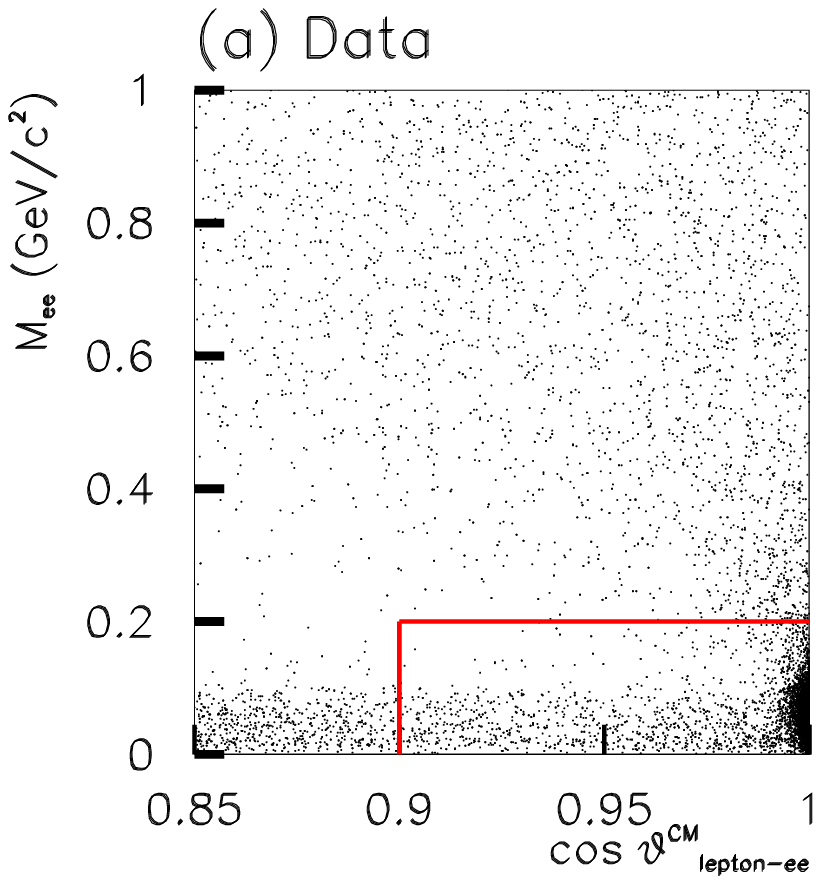}}
\hspace*{-1.2cm}
 \resizebox{0.37\textwidth}{0.37\textwidth}{\includegraphics
{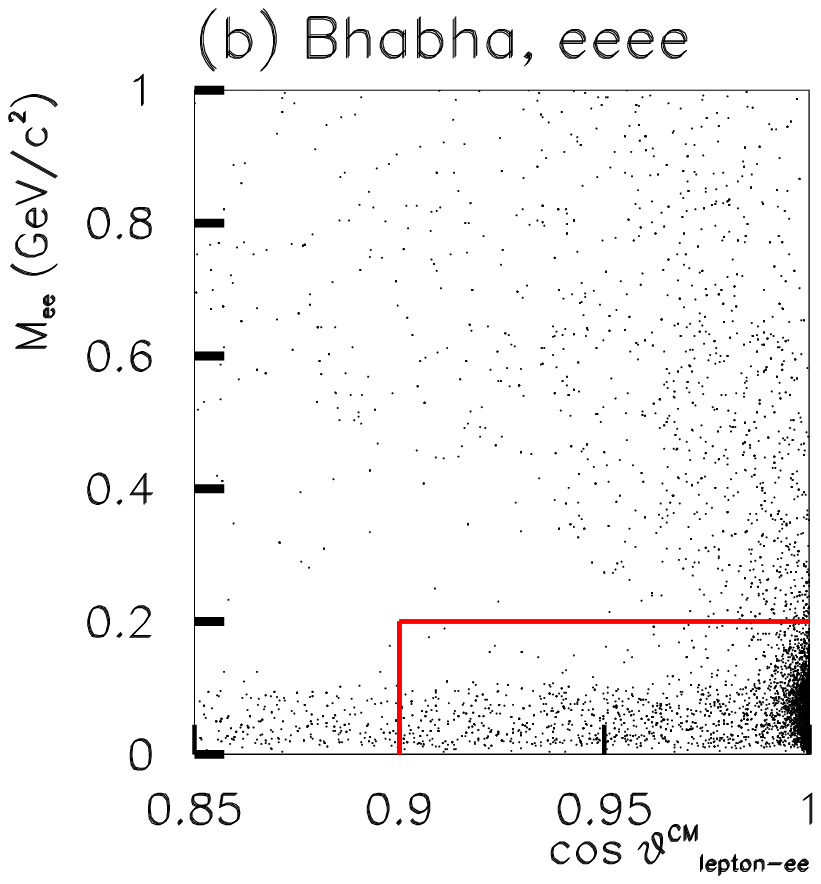}}
\hspace*{-1.2cm}
 \resizebox{0.37\textwidth}{0.37\textwidth}{\includegraphics
{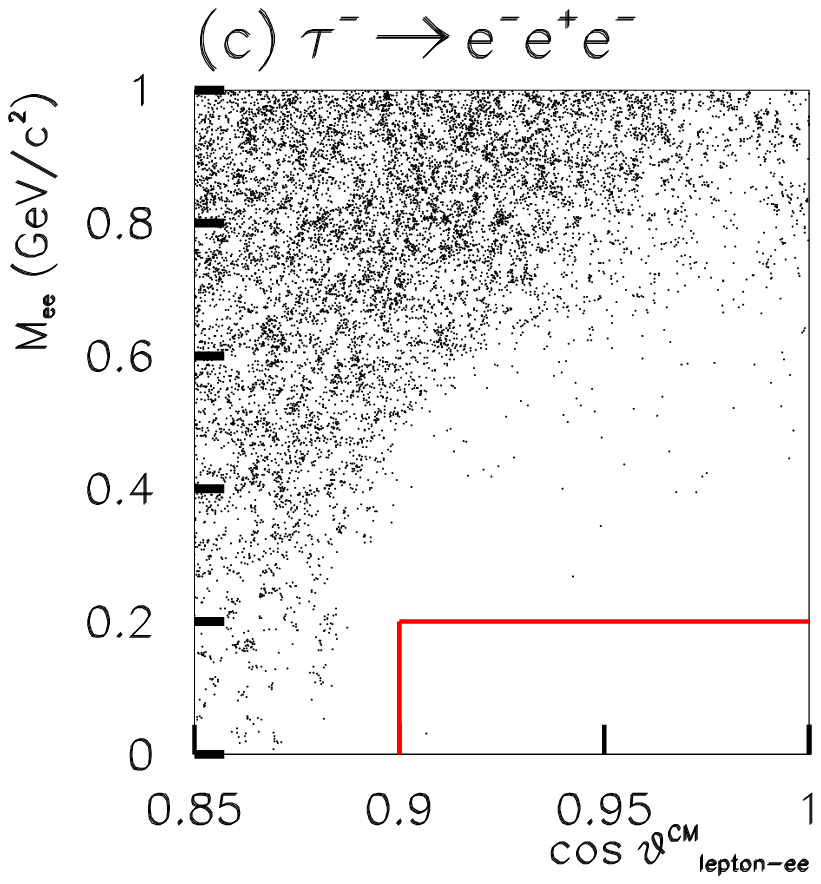}}
 \vspace*{-0.5cm}
\caption{
Scatter-plots
of the
{reconstructed  invariant mass} 
of the $e^+e^-$ pair
($M_{ee}$) vs.  cosine of the  opening angle between
the direction of the $e^+e^-$ pair and the 
other electron 
($\cos \theta^{\rm CM}_{\rm{lepton}-ee}$)
for
(a) data, (b) Bhabha and $eeee$,
(c) signal MC ($\tau^-\to e^-e^+e^-$)
}
\label{fig:convsersion}
\end{figure}

Conversions ($\gamma\to e^+e^-$) 
are a large background for the 
{$\tau^-\-\to e^-e^+e^-$} and $\mu^-e^+e^-$ modes.
We require that 
the cosine of the opening angle between 
the direction of the $e^+e^-$ pair 
and the other lepton
in the {CM} system,
($\cos \theta^{\rm CM}_{\rm{lepton}-ee}$)
be less than 0.90
if the invariant mass of the $e^+e^-$ pair
($M_{ee}$) is less than 0.2 GeV/$c^2$ 
for these modes.
As shown in Fig.~\ref{fig:convsersion} 
{for the $\tau^-\to e^-e^+e^-$ mode 
(two entries from each event),} 
the signal efficiency is not affected by this cut, while
the large background from the conversions can be reduced.

%
%
For the $\tau^-\to e^-e^+e^-$ and
$\tau^-\to e^-\mu^+\mu^-$ modes,
the charged track on the tag side is required not to be an electron 
by applying 
${\cal P}(e)<0.1$ since a large background still remains from two-photon
and Bhabha events.
Furthermore,
we reject the event if the charged track on the tag side 
is in 
gaps between barrel and endcap of
the ECL.
%
%
To reduce 
{backgrounds} from Bhabha and $\mu^+\mu^-$ events,
we require that the momentum in the CM system of 
the charged track on the tag side be less than 4.5 {GeV/$c$}
for the $\tau^-\to e^-e^+e^-$ and $\tau^-\to\mu^-e^+e^-$ modes.

\begin{figure}
\begin{center}
 \resizebox{0.65\textwidth}{0.65\textwidth}{\includegraphics
 {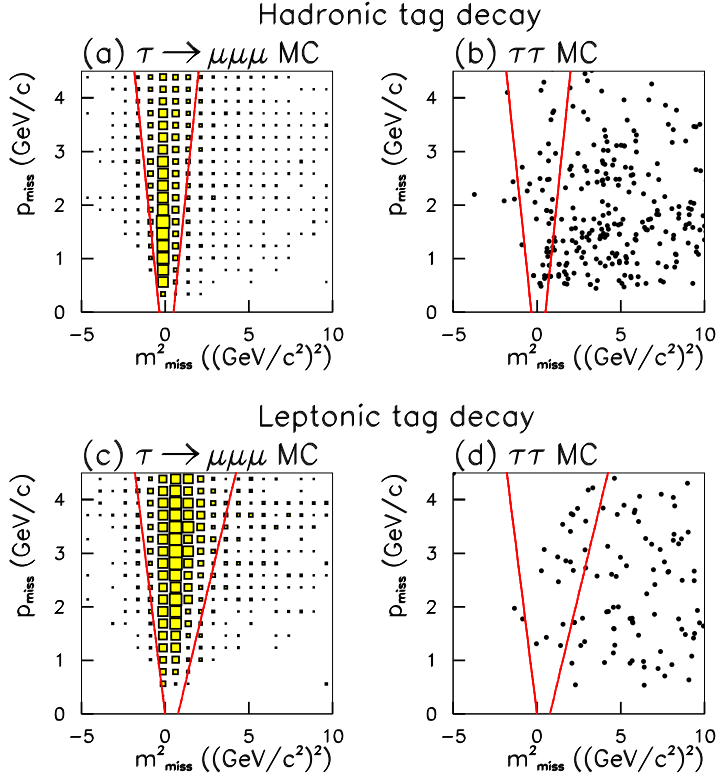}}
 \vspace*{-0.5cm}
 \caption{
Scatter-plots of
{$p_{\rm miss}$ 
{vs.} 
$m_{\rm miss}^2$:
(a) and (b)
show
the signal MC 
($\tau^-\to\mu^-\mu^+\mu^-$)
and 
the generic $\tau^+\tau^-$ MC 
distributions,
respectively,
for the hadronic tag
while (c) and
(d) show
the same distributions
for the leptonic one.}
Selected regions are indicated by lines.}
\label{fig:pmiss_vs_mmiss2}
\end{center}
\end{figure}

%
%
Finally, 
to suppress  backgrounds from generic 
$\tau^+\tau^-$ and $q\bar{q}$ events, 
we apply a selection based on the magnitude of the missing momentum ${p}_{\rm{miss}}$ 
and missing mass squared $m^2_{\rm{miss}}$ for all modes 
except for 
$\tau^-\to e^+\mu^-\mu^-$ and $\mu^+ e^-e^-$.
We do not apply this cut 
for the latter modes
since  {backgrounds  for them are} 
much 
{smaller.}
We apply different selection criteria depending on 
whether the $\tau$ decay on the tag side is  hadronic or leptonic:
{two} neutrinos are 
{emitted}
if the $\tau$ decay {on} the tag side is
{leptonic,}
while
one neutrino is 
{emitted}
if the $\tau$ decay {on} the tag
side is {a hadronic one.}
Therefore,
we separate events into
two classes
according to the track {on} the tag side: leptonic or hadronic.
The selection criteria are listed in Table~\ref{tbl:misscut};  
the distributions of $m^2_{\rm{miss}}$ and $p_{\rm{miss}}$ for 
hadronic and leptonic decays are shown in Fig.~\ref{fig:pmiss_vs_mmiss2}. 
\begin{table}
\begin{center}
\caption{
The selection criteria for 
the missing momentum ($p_{\rm{miss}}$) and
missing mass squared ($m^2_{\rm miss}$) 
correlations for each mode,
$p_{{\rm miss}}$ is in GeV/$c$
and  $m^2_{\rm miss}$ is in $({\rm{GeV}}/c^2)^2$.
}
\label{tbl:misscut}
\begin{tabular}{c|cc} \hline\hline
Mode & Hadronic tag mode & Leptonic tag mode \\ \hline
$\tau^-\to\mu^-\mu^+\mu^-$ & $p_{\rm miss} > -3.0m^2_{\rm miss}-1.0$ &
 $p_{\rm miss} > -2.5m^2_{\rm miss}$ \\
$\tau^-\to \mu^- e^+ e^-$ &  $p_{\rm miss} > 3.0m^2_{\rm miss}-1.5$  &
 $p_{\rm miss} > 1.3m^2_{\rm miss}-1$ \\
$\tau^-\to e^- \mu^+ \mu^-$ & & \\ \hline
$\tau^-\to e^-e^+ e^-$ & $p_{\rm miss} > -3.0m^2_{\rm miss}-1.0$ &
 $p_{\rm miss} > -2.5m^2_{\rm miss}$ \\
 &  $p_{\rm miss} > 4.2m^2_{\rm miss}-1.5$  &
 $p_{\rm miss} > 2.0m^2_{\rm miss}-1$ \\ \hline
$\tau^-\to e^+ \mu^- \mu^-$ & N.A.  & N.A. \\
$\tau^-\to \mu^+ e^- e^-$ &    & \\ \hline\hline
\end{tabular}
\end{center}
\end{table}

\section{Signal and Background Estimation}

The signal candidates are examined in the two-dimensional plots 
of the $\ell^-\ell^+\ell^-$ invariant
mass~($M_{\rm {3\ell}}$), and 
the difference of their energy from the 
beam energy in the CM system~($\Delta E$).
A signal event should have $M_{\rm {3\ell}}$
close to the $\tau$-lepton mass and
$\Delta E$ close to zero.
For all modes,
the $M_{\rm {3\ell}}$ and $\Delta E$  resolutions are parameterized
from fits to the signal MC distributions  
with  an asymmetric Gaussian function that takes into account 
initial state radiation.
The resolutions of 
$M_{\rm 3\ell}$ and $\Delta E$ for each mode are summarized 
in Table~\ref{tbl:reso_del_e_m}.
\begin{table}
\begin{center}
\caption{Summary {of} $M_{\rm 3\ell}$  and 
$\Delta E$ 
{resolutions}}
\label{tbl:reso_del_e_m}
\begin{tabular}{c|cccc} \hline\hline 
Mode
& $\sigma^{\rm{high}}_{M_{\rm{3\ell}}}$ (MeV/$c^2$)  
& $\sigma^{\rm{low}}_{M_{\rm{3\ell}}}$ (MeV/$c^2$)
& $\sigma^{\rm{high}}_{\Delta E}$ (MeV)      
&  $\sigma^{\rm{low}}_{\Delta E}$ (MeV)
 \\ \hline
$\tau^-\to\mu^-\mu^+\mu^-$
& 4.8  &  5.4 & 12.5 & 15.7 \\
$\tau^-\to e^- e^+ e^-$
& 5.1 & 7.8 & 13.4 & 25.1 \\
$\tau^-\to e^-\mu^+\mu^-$
& 5.1  &  5.6 & 12.1 & 19.6  \\
$\tau^-\to\mu^- e^+ e^-$
 & 5.0  &  6.6 & 13.4 & 21.3  \\
$\tau^-\to e^+\mu^-\mu^-$
& 5.0  &  6.0 & 13.3 & 19.9  \\
 $\tau^-\to\mu^+ e^- e^-$
 & 5.4  &  6.7 & 13.8 & 23.0  \\  \hline\hline
\end{tabular}
\end{center}
\end{table}

To evaluate the branching fractions, 
we define  elliptical signal regions determined 
from fits  to the $M_{\rm {3\ell}}$ and $\Delta E$ distributions 
as shown in Table~\ref{tbl:reso_del_e_m}.
These signal regions are optimized by signal MC, 
so that 90\% of the signal events that passed  all the
selections are contained in the signal region. 

\begin{figure}
\begin{center}
\resizebox{0.4\textwidth}{0.4\textwidth}{\includegraphics
{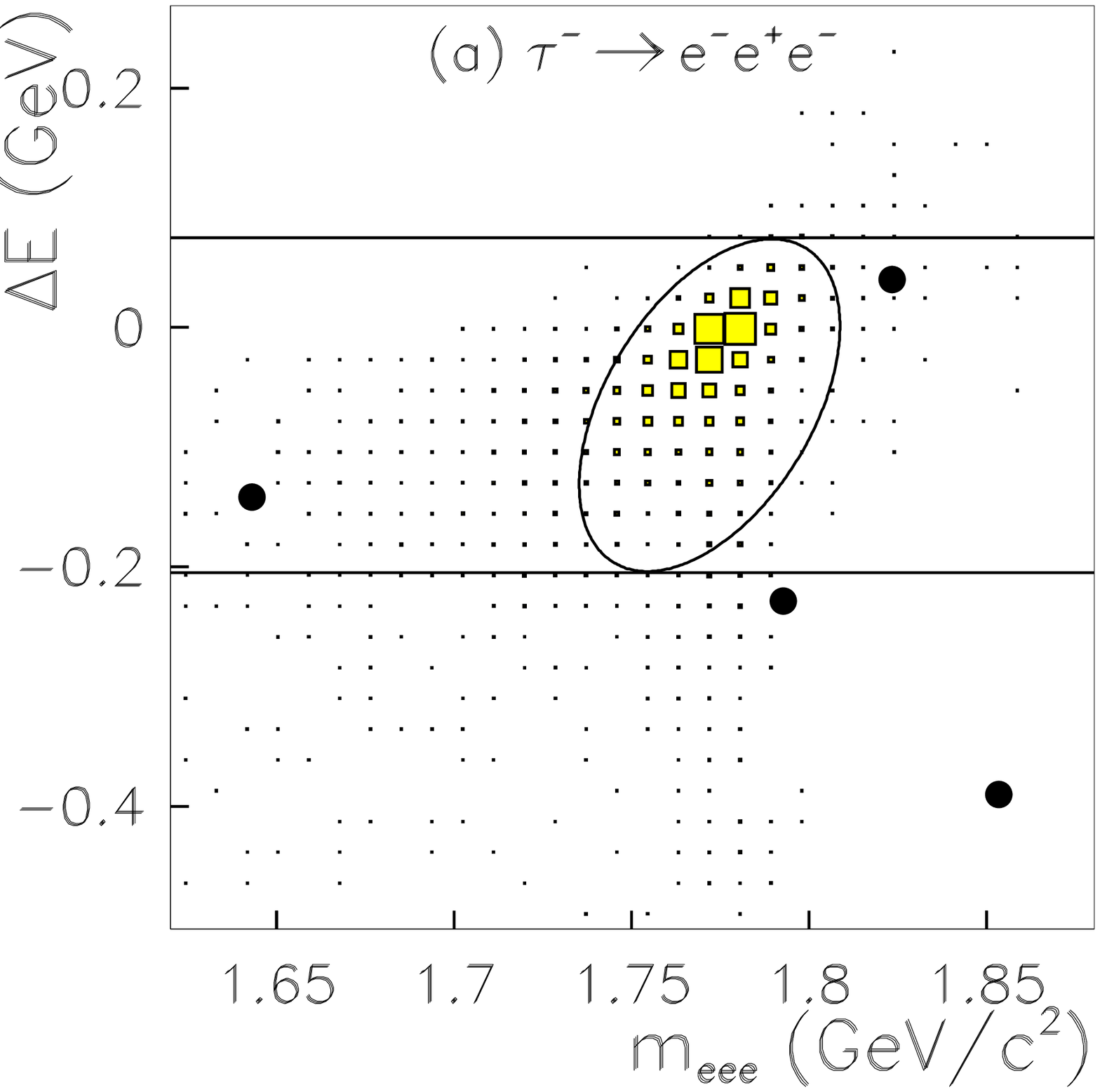}}
\resizebox{0.4\textwidth}{0.4\textwidth}{\includegraphics
{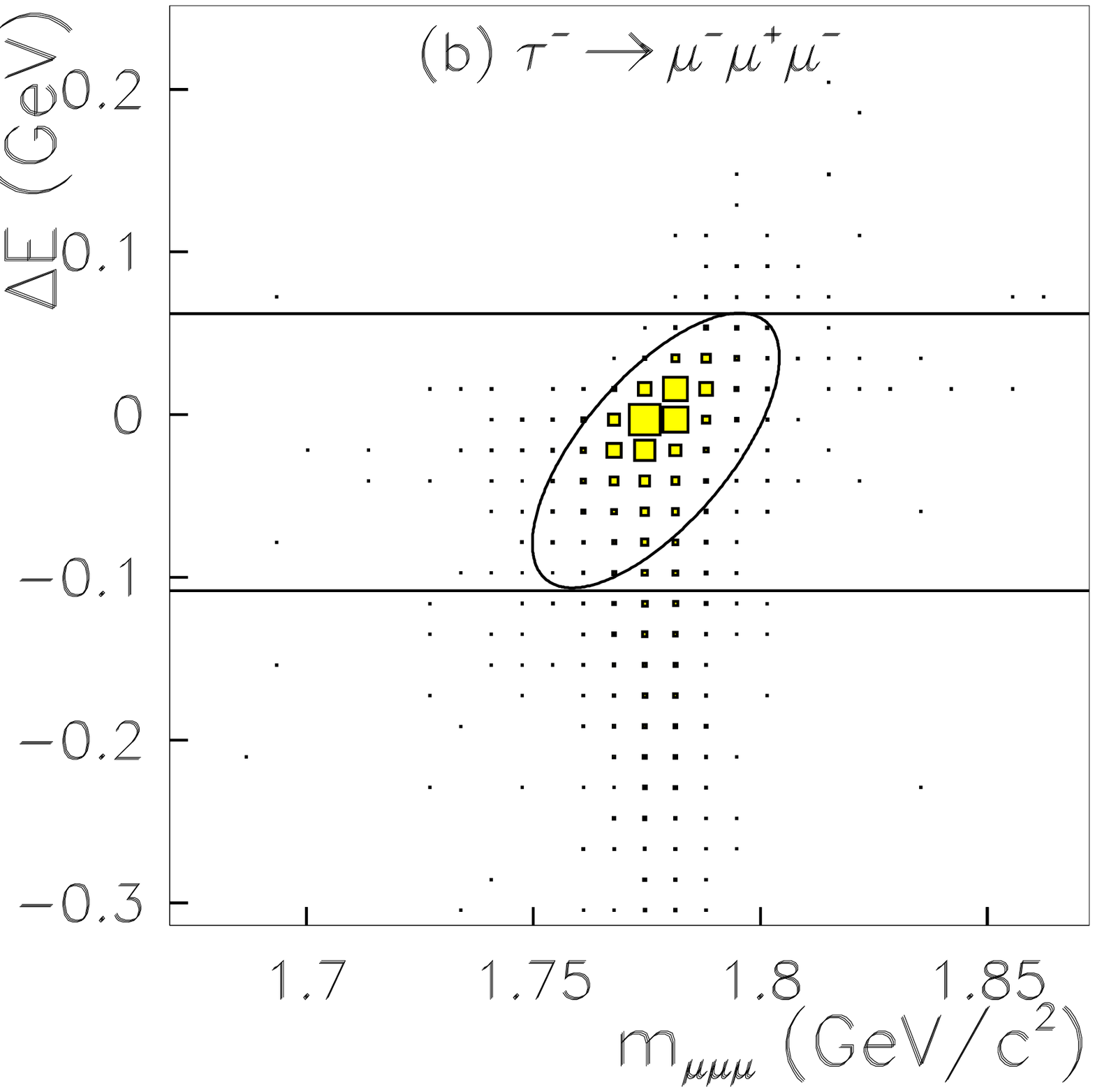}}\\
\vspace*{-0.5cm}
\resizebox{0.4\textwidth}{0.4\textwidth}{\includegraphics
{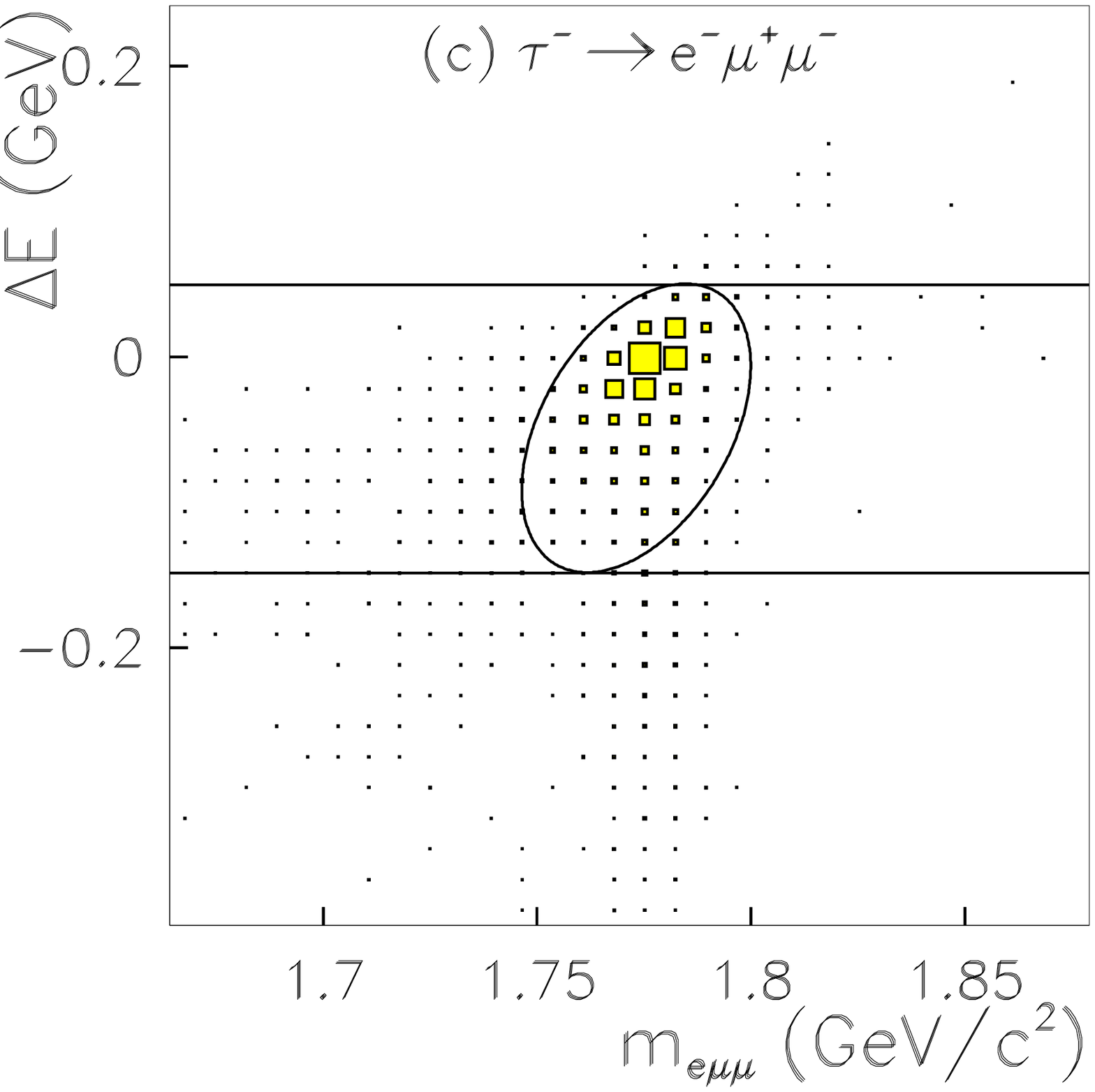}}
\resizebox{0.4\textwidth}{0.4\textwidth}{\includegraphics
{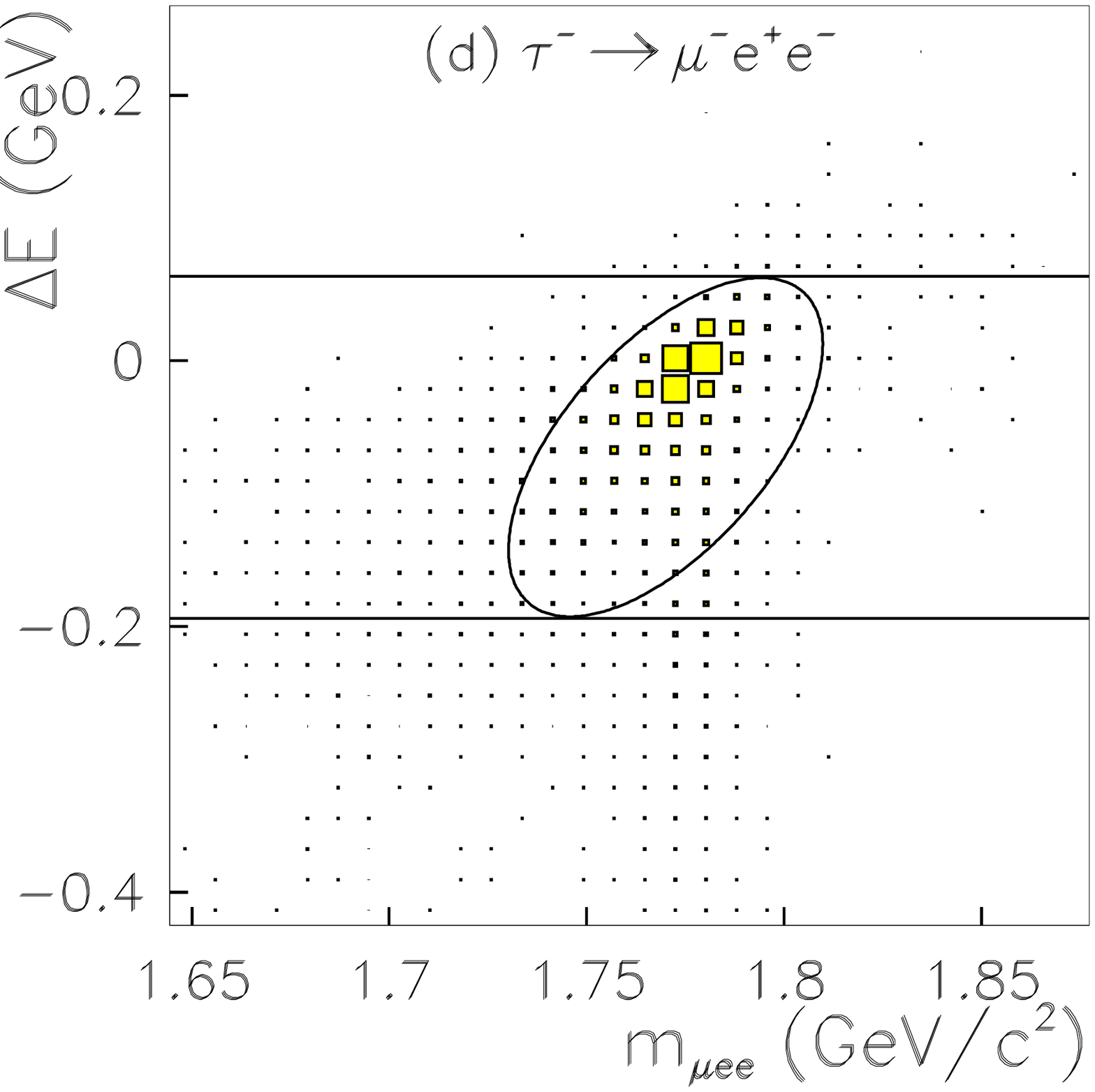}}\\
\vspace*{-0.5cm}
\resizebox{0.4\textwidth}{0.4\textwidth}{\includegraphics
{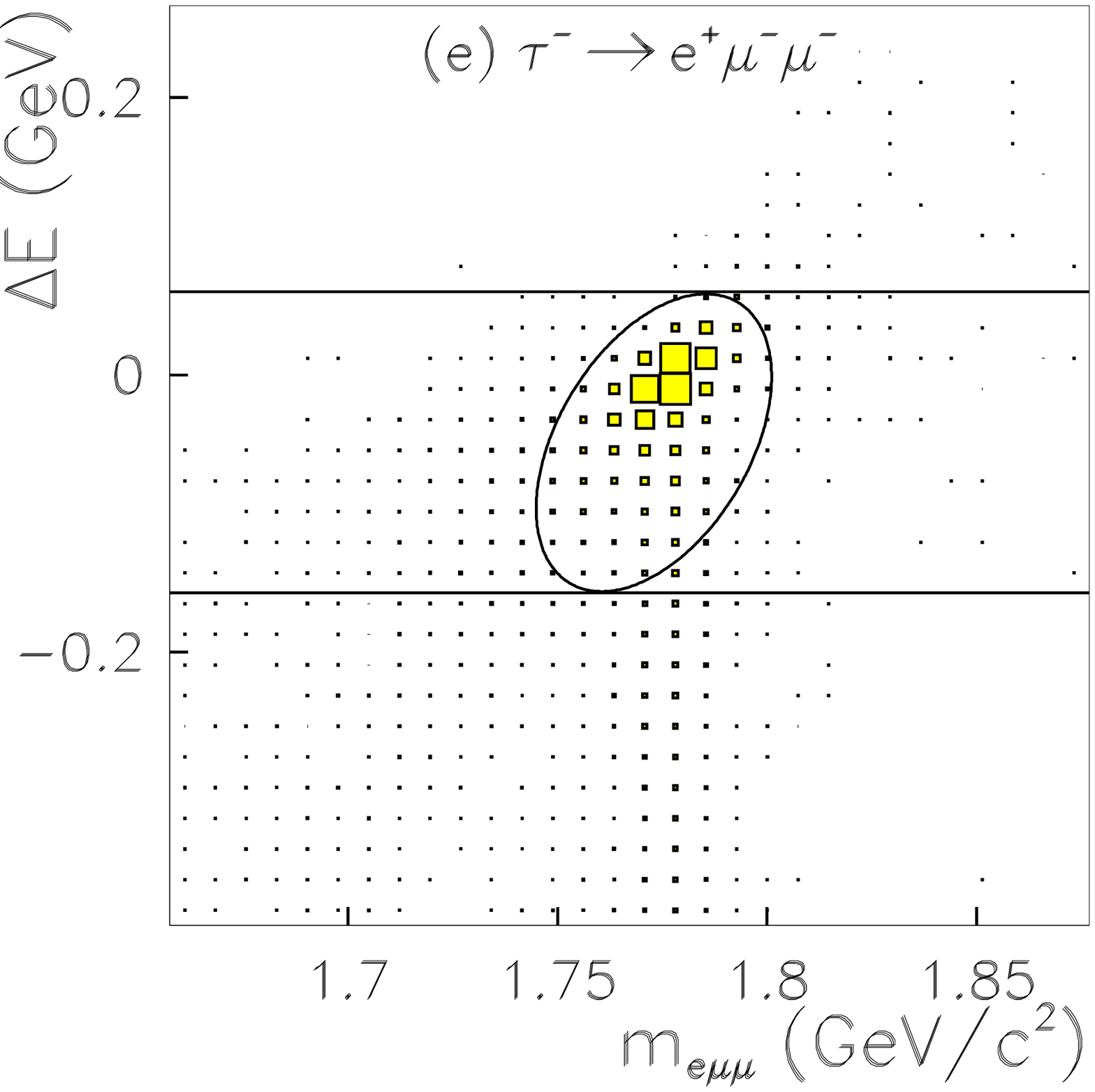}}
\resizebox{0.4\textwidth}{0.4\textwidth}{\includegraphics
{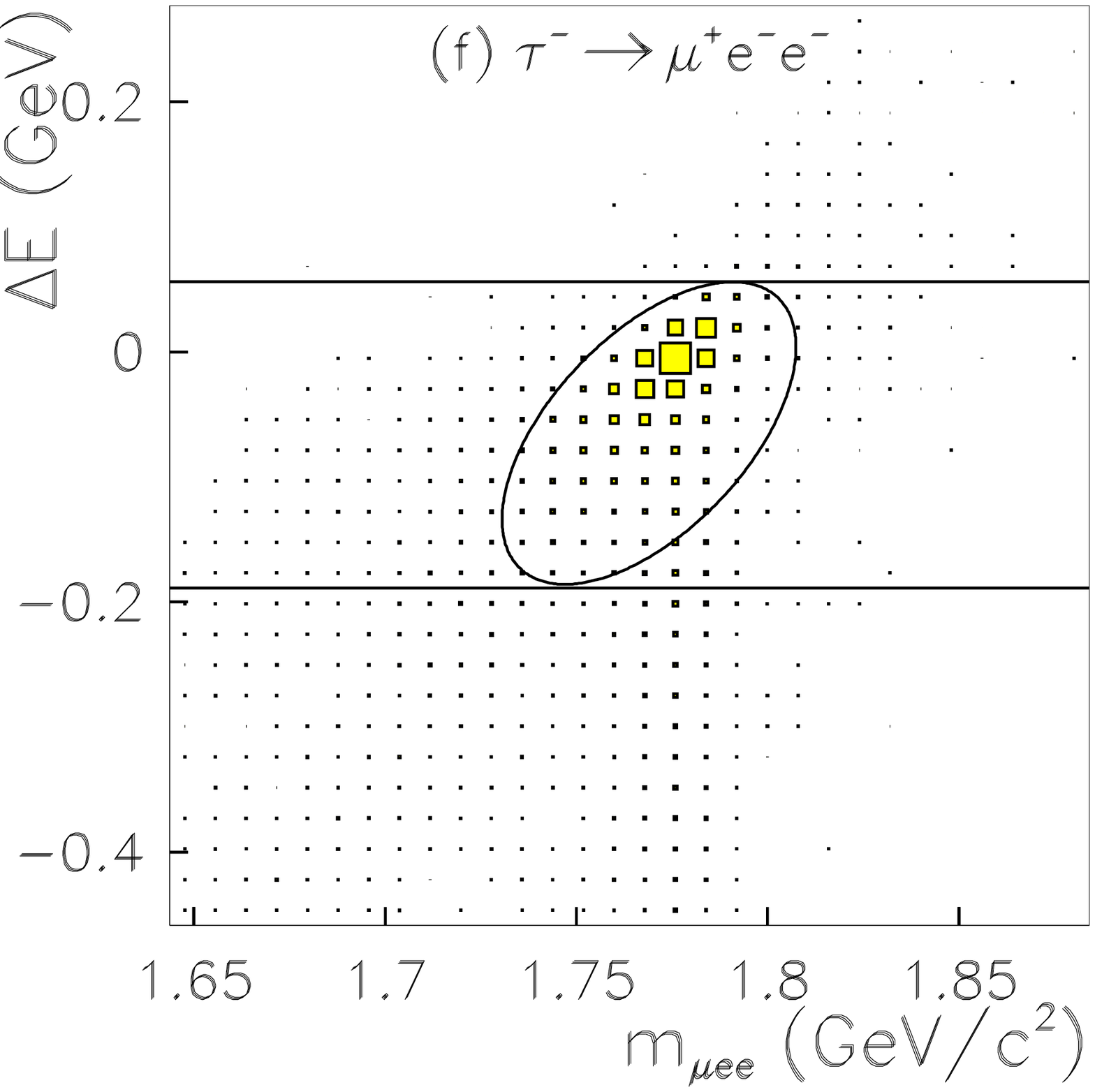}}
\caption{
Scatter-plots in the
$M_{3\ell}$ -- $\Delta{E}$ plane:
(a), (b), (c) , (d), (e) and (f) correspond to
the $\pm 20 \sigma$ area for
the
$\tau^-\rightarrow e^-e^+e^-$,
$\tau^-\rightarrow\mu^-\mu^+\mu^-$,
$\tau^-\rightarrow e^-\mu^+\mu^-$,
$\tau^-\rightarrow\mu^- e^+e^-$,
$\tau^-\rightarrow e^+\mu^-\mu^-$ and
$\tau^-\rightarrow \mu^+e^-e^-$
modes, respectively.
The data are indicated by the solid circles.
The filled boxes show the MC signal distribution
with arbitrary normalization.
The elliptical signal 
{regions} 
shown by a solid curve 
are used for evaluating the signal yield.
The region between the horizontal solid lines excluding
the signal region is
used to estimate the expected background in the elliptical region. 
}
\label{fig:openbox}
\end{center}
\end{figure}

We blind the data in the signal region and estimate 
the signal efficiency and the number of the backgrounds from the MC 
and the data outside the signal region, 
so as not to bias our choice of selection criteria. 
Figure~\ref{fig:openbox} shows scatter-plots
for the data and the signal MC distributed over $\pm 20\sigma$
in the $M_{\rm{3\ell}}-\Delta E$ plane.
No events are observed 
{outside} 
the signal region for any modes except 
for 
$\tau^-\to e^-e^+e^-$ in which four events are  found.
{The remaining background events  
in the  $\tau^-\to e^-e^+e^-$ mode
are expected to come from
a Bhabha  electron or $\tau^-\to e^-\nu_{\tau}\bar{\nu_{e}}$ and 
two electrons from a gamma conversion.}
{The final estimate of 
the number of the background events 
is based on the data 
{with looser selection criteria}
in the $M_{\rm{3\ell}}$ 
sideband region,} 
which is defined as the box 
inside
{the} horizontal lines {but}
excluding the signal region,
as shown by the lines in Fig.~\ref{fig:openbox}.
Assuming that the background distribution is uniform in the sideband region, 
the number of background events in the signal box is estimated by 
interpolating the number of observed events in the sideband {region} 
into the signal {region.}
The signal efficiency and the number of expected background 
events for each mode 
are summarized in Table~\ref{tbl:eff}.

We estimate the  systematic uncertainties
due to the lepton identification, the charged track finding, the MC statistics, and  the integrated luminosity. 
The uncertainty due to the trigger efficiency is negligible compared with the other uncertainties.
The uncertainties due to the lepton {identification} 
{are} 
2.2\% per each electron and 2.0\% per each muon.
The uncertainty due to the charged track finding is estimated to be 1.0\% per charged track.
The uncertainty due to the $e$-veto on the tag side applied for 
{the} 
$\tau^-\to e^-e^+e^-$ and $\tau^-\to e^-\mu^+\mu^-$ modes is estimated to be the same as  
the uncertainty due to the electron identification.
The uncertainties due to MC statistics and luminosity
are estimated to be 
{(0.5 - 0.9)\%} and 1.4\%, respectively.
All these uncertainties are added in quadrature, 
and the total systematic uncertainty for each mode is listed in Table~\ref{tbl:eff}.
\begin{table}
\begin{center}
\caption{ The signal efficiency($\varepsilon$), 
the number of the expected background {events}  ($N_{\rm BG}$)
estimated from the  sideband data, 
{total} 
systematic uncertainty  ($\sigma_{\rm syst}$),
{the} number of the observed events 
in the signal region ($N_{\rm obs}$), 
90\% C.L. upper limit on the number of signal events including 
systematic uncertainties~($s_{90}$) 
and 90\% C.L. upper limit on the branching 
fraction~($\cal{B}$)
for each individual mode. }
\label{tbl:eff}
\begin{tabular}{c|cccccc}\hline \hline
Mode &  $\varepsilon$~{(\%)} & 
$N_{\rm BG}$  & $\sigma_{\rm syst}$~{(\%)}
& $N_{\rm obs}$ & $s_{90}$ & 
${\cal{B}}(\times10^{-8})$ \\ \hline
$\tau^-\to e^-e^+e^-$ &  6.00 & 
 0.40$\pm$0.30 & 9.8 &
 0 & 2.10 & 3.6 \\ 
$\tau^-\to\mu^-\mu^+\mu^-$ & 7.64 & 0.07$\pm{0.05}$ & 
 7.4 &
0 &  2.41 & 3.2 \\
$\tau^-\to e^-\mu^+\mu^-$ &  6.08 & 0.05$\pm{0.03}$ 
  & 9.5 &
 0 & 2.44 & 4.1\\
$\tau^-\to \mu^-e^+e^-$ &  9.29 & 0.04$\pm{0.04}$
  & 7.8 &
0 & 2.43  & 2.7\\ 
$\tau^-\to e^+\mu^-\mu^-$ &  10.8 & 0.02$\pm{0.02}$
 & 7.6 & 
0 &  2.44& 2.3\\
$\tau^-\to \mu^+e^-e^-$ &  12.5 &
0.01$\pm{0.01}$
 & 7.7 & 
0 & 2.46  & 2.0\\ 
\hline\hline
\end{tabular}
\end{center}
\end{table}

\section{Upper Limits on the branching fractions}

Finally, we open the blind and 
{find no events} in the signal region.
Since 
no events are
{observed in the signal region,}
we set upper limits on the branching fractions 
of $\tau^-\to\ell^-\ell^+\ell^-$.
The 90\% C.L. upper limit on the number of the signal events 
including  a systematic uncertainty~($s_{90}$) is obtained 
from the number of expected {background events} and observed data, 
calculated by the POLE program without conditioning \cite{pole}, which 
is based on the Feldman-Cousins method~\cite{cite:FC}.
The upper limit on the branching fraction ($\cal{B}$) is then given by
\begin{equation}
{{\cal{B}}(\tau^-\to\ell^-\ell^+\ell^-) <
\displaystyle{\frac{s_{90}}{2N_{\tau\tau}\varepsilon{}}},}
\end{equation}
where $N_{\tau\tau}$ is the number of $\tau^+\tau^-$pairs, and 
$\varepsilon$ is the signal efficiency.
{$N_{\tau\tau} =  492\times 10^6$} is obtained 
from 535 fb${}^{-1}$ of integrated luminosity times 
the cross section of the $\tau$ pair production, which 
is calculated 
{in the updated version of 
KKMC~\cite{tautaucs} to be 
$\sigma_{\tau\tau} = 0.919 \pm 0.003$ nb.}
The 90\% C.L. upper limits on the branching fractions 
${\cal{B}}(\tau^-\rightarrow \ell^- \ell^+\ell^-)$  are in the range
between 
{$2.0 \times 10^{-8}$ and $4.1 \times 10^{-8}$} 
and are summarized in Table~\ref{tbl:eff}.
These results improve  the  {best} 
previously published upper limits~\cite{cite:3l_belle,cite:3l_babar} 
by factors from {4.9 to 7.0.}

%
%
%

\section{Summary}
We have searched for {lepton-flavor-violating} $\tau$ decays 
into three leptons using 535 fb$^{-1}$ of data.
No events 
are observed and
we set the 90\% C.L. upper limits 
{on the branching fractions:} 
${\cal{B}}(\tau^-\rightarrow e^-e^+e^-) < 3.6\times 10^{-8}$, 
${\cal{B}}(\tau^-\rightarrow \mu^-\mu^+\mu^-) < 3.2\times 10^{-8}$, 
${\cal{B}}(\tau^-\rightarrow e^-\mu^+\mu^-) < 4.1\times 10^{-8}$, 
${\cal{B}}(\tau^-\rightarrow \mu^-e^+e^-) < 2.7\times 10^{-8}$, 
${\cal{B}}(\tau^-\rightarrow e^+\mu^-\mu^-) < 2.3\times 10^{-8}$
and  
${\cal{B}}(\tau^-\rightarrow \mu^+e^-e^-) < 2.0\times 10^{-8}$.
These results improve the {best} 
previously published upper limits
by factors from 4.9 to 7.0.
These more stringent upper limits can be used
to constrain the space of parameters in various models beyond
the SM.


\section*{Acknowledgments}

The authors are grateful to A.~Buras and Th.~Mannel for fruitful 
discussions.
We thank the KEKB group for the excellent operation of the
accelerator, the KEK cryogenics group for the efficient
operation of the solenoid, and the KEK computer group and
the National Institute of Informatics for valuable computing
and Super-SINET network support. We acknowledge support from
the Ministry of Education, Culture, Sports, Science, and
Technology of Japan and the Japan Society for the Promotion
of Science; the Australian Research Council and the
Australian Department of Education, Science and Training;
the National Science Foundation of China and the Knowledge
Innovation Program of the Chinese Academy of Sciences under
contract No.~10575109 and IHEP-U-503; the Department of
Science and Technology of India; 
the BK21 program of the Ministry of Education of Korea, 
the CHEP SRC program and Basic Research program 
(grant No.~R01-2005-000-10089-0) of the Korea Science and
Engineering Foundation, and the Pure Basic Research Group 
program of the Korea Research Foundation; 
the Polish State Committee for Scientific Research; 
the Ministry of Education and Science of the Russian
Federation and the Russian Federal Agency for Atomic Energy;
the Slovenian Research Agency;  the Swiss
National Science Foundation; the National Science Council
and the Ministry of Education of Taiwan; and the U.S.\
Department of Energy.


%
%
%

\end{document}